\documentclass[twocolumn,preprintnumbers,amsmath,amssymb]{revtex4}

\usepackage{mathrsfs}
\usepackage{graphicx}
\usepackage{dcolumn}
\usepackage{bm}
\usepackage{amsmath}

\begin{document}

\title{Minimal Energy Cost of Thermodynamic Information Processing: Measurement and Information Erasure}
\author{Takahiro Sagawa$^1$}
\author{Masahito Ueda$^{1,2}$}
\affiliation{$^1$Department of Physics, University of Tokyo,
7-3-1 Hongo, Bunkyo-ku, Tokyo, 113-8654, Japan \\
$^2$ERATO Macroscopic Quantum Control Project, JST, 2-11-16 Yayoi, Bunkyo-ku, Tokyo 113-8656, Japan
}
\date{\today}

\pacs{05.70.Ln, 05.30.-d, 03.67.-a}

\begin{abstract}
The fundamental lower bounds on the thermodynamic energy cost of measurement and information erasure are determined. The lower bound on the erasure validates Landauer's principle for a symmetric memory; for other cases, the bound indicates the breakdown of the principle.
Our results  constitute the second law of ``information  thermodynamics,'' in which information content and  thermodynamic variables are treated on an equal footing.
\end{abstract}

\maketitle

The fundamental lower bound on the thermodynamic energy cost of information processing has been a topic of active research~\cite{Demon,Szilard,Brillouin,Landauer,Bennett2,Piechocinska,Barkeshli,Maroney,Sagawa-Ueda,Demons}.
According to Landauer's principle~\cite{Landauer}, on average, at least $k_{\rm B} T \ln 2$ of work is required to erase one bit of information from a memory.
Recent developments in nanoscience have enabled the direct measurement of such minuscule amounts of work for small nonequilibrium thermodynamic systems~\cite{Nonequilibrium}.  
Moreover, advances in molecular devices~\cite{Aviram2} and nanomachines~\cite{Serreli} have necessitated a deeper understanding of thermodynamic information processing.  
In view of these developments, it is essential to identify the fundamental lower bound on the thermodynamic energy cost of information processing.
In this Letter, we derive the minimum work that must be performed on a memory for measurement and information erasure.  
Our results are proved rigorously for  classical information processing and proved for quantum information processing after making an additional assumption .
Our results are independent of the detailed characteristics of the system and memory, and therefore, they can potentially be applied in many areas of information processing. 

We consider a memory M that  stores information on  the outcome of a measurement.  
While we formulate M as a quantum system, our formulation is also valid in the classical limit.
Let $\textbf H^{\rm M}$ be the Hilbert space of M. 
We decompose $\textbf H^{\rm M}$ into mutually orthogonal subspaces $\textbf H_k^{\rm M}$ ($k = 1, 2, \cdots, N$), where the $k$'s describe the measurement outcomes:  $\textbf H^{\rm M} = \bigoplus_k \textbf H_k^{\rm M}$.  
We assume that outcome ``$k$'' is stored in M if the support of the density operator of the memory belongs to $\textbf H^{\rm M}_k$.  
Without loss of generality, we assume that  $k=0$ corresponds to the standard state of M.  
The Hamiltonian of M corresponding to outcome ``$k$'' is denoted by $\hat H^{\rm M}_k \equiv \sum_{i} \varepsilon_{ki} | \varepsilon_{ki} \rangle \langle \varepsilon_{ki} |$, where $\{ | \varepsilon_{ki} \rangle \}_{i}$ is an orthonormal basis set of $\textbf H_k^{\rm M}$.  

Suppose that we perform a measurement on a thermodynamic system S by an isothermal process at temperature $T$ and find outcome ``$k$'' with probability $p_k$.  The information on this outcome is stored in  memory M.  
We note that $p_k$ depends on the state of the measured system S but is independent of the structure of memory M.
We assume that  M is in contact with a heat bath B at temperature $T$. 
The total Hamiltonian is given by $\hat H^{\rm MB} (t) =  \hat H^{\rm M}(t) + \hat H^{\rm int} (t) + \hat H^{\rm B}$, where $\hat H^{\rm M} \equiv \bigoplus_k \hat H_k^{\rm M}$, and $\hat H^{\rm int} (t)$ is the interaction Hamiltonian between M and B~\cite{Sagawa-Ueda}.
We consider the measurement process from $t= 0$ to $t = \tau$ and assume that $\hat H^{\rm M}(0) = \hat H^{\rm M} (\tau ) = \hat H^{\rm M}$ and $\hat H^{\rm int} (0) = \hat H^{\rm int} (\tau ) = 0$.

The initial state of  M is assumed to obey the canonical distribution at temperature $T$ subject to the constraint $k=0$.  System S is initially separated from  M and  B.  The total density operator is given by $\hat \rho_{\rm i}^{\rm SMB} = \hat \rho_{\rm i}^{\rm S} \otimes \hat \rho_{0,{\rm can}}^{\rm M} \otimes \hat \rho_{\rm can}^{\rm B}$,
where $\hat \rho_{0, {\rm can}}^{\rm M} \equiv \exp (- \beta \hat H_{0}^{\rm M})/Z_0^{\rm M}$ with $Z_0^{\rm M} \equiv {\rm tr} [\exp (- \beta \hat H_0^{\rm M})]$ and $\hat \rho_{\rm can}^{\rm B} \equiv \exp (- \beta \hat H^{\rm B}) / Z^{\rm B}$ with $Z^{\rm B} \equiv {\rm tr} [\exp (- \beta \hat H^{\rm B})]$. We do not assume that the initial state of S is in thermodynamic equilibrium.  

Next we perform a measurement on S.  First, M unitarily interacts with system S according to the unitary operator $\hat U_{\rm int}$.  By this interaction, M becomes entangled with  system S.  The state of  M is then measured and projected onto the subspace corresponding to the measurement outcome ``$k$.''  The latter process is described by the projection operator $\hat P^{\rm M}_k \equiv \sum_{i} | \varepsilon_{ki} \rangle \langle \varepsilon_{ki} |$.  
Immediately  after the measurement, the total density operator is given by $\hat \rho'^{\rm SMB} =  \sum_k  \hat P^{\rm M}_k \hat U_{\rm int} \hat \rho_{\rm i}^{\rm SMB} \hat U_{\rm int}^\dagger \hat P^{\rm M}_k$. 
We assume that  
\begin{equation}
\hat \rho'^{\rm SMB} = \sum_{k, i}   \hat M_{ki} \hat \rho_{\rm i}^{\rm S} \hat M_{ki}^\dagger   \otimes \hat \rho_{ki}^{\rm MB},
\label{assumption}
\end{equation}
where $\hat M_{ki}$'s are the measurement operators that give the positive operator-valued measure (POVM) $\hat E_k \equiv \sum_{i} \hat M_{ki}^\dagger \hat M_{ki}$~\cite{Davies-Lewis-Kraus}, and $\hat \rho_{ki}^{\rm MB}$'s are the density operators of M and B that are mutually orthogonal.  The assumption in Eq.~(\ref{assumption}) can be justified for a classical case, as shown later.  Our results are also applicable to quantum systems that satisfy Eq.~(\ref{assumption}). We note that $p_k = {\rm tr}(\hat E_k \hat \rho_{\rm i}^{\rm S}) = \sum_{i}p_{ki}$ with $p_{ki} \equiv {\rm tr}(\hat M_{ki}^\dagger \hat M_{ki} \hat \rho_{\rm i}^{\rm S})$.
Finally, S is detached  from  M and  B, and then, M+B unitarily evolves according to the unitary operator $\hat U_{\rm f}$. The final state is given by $\hat \rho_{\rm f}^{\rm SMB} =  \sum_{k, i}  \hat M_{ki} \hat \rho_{\rm i}^{\rm S} \hat M_{ki}^\dagger  \otimes \hat U_{\rm f} \hat \rho_{ki}^{\rm MB} \hat U^{\dagger}_{\rm f}$.

Let $F_{k}^{\rm M} \equiv - k_{\rm B}T \ln Z_k^{\rm M}$ with $Z_k^{\rm M} \equiv {\rm tr} [\exp ( - \beta \hat H^{\rm M}_k)]$ be the Helmholtz free energy of M with measurement outcome ``$k$,'' and let $\hat \rho_k^{\rm MB} \equiv \sum_{i} (p_{ki}/p_k) \hat \rho_{ki}^{\rm MB}$ be the post-measurement state of M with outcome $k$. We define the average change in the free energy due to the measurement as $\Delta F^{\rm M} \equiv \sum_k p_k F_k^{\rm M} - F_{0}^{\rm M}$ and the ensemble average of work performed on M during the measurement as $W_{\rm meas}^{\rm M} \equiv \sum_k p_k [ {\rm tr}(\hat \rho_k^{\rm MB} \hat H^{\rm M}_k) + {\rm tr}(\hat \rho_k^{\rm MB} \hat H^{\rm B}) ] - [{\rm tr}(\hat \rho_{0, {\rm can}}^{\rm M} \hat H^{\rm M}_0) + {\rm tr}(\hat \rho_{\rm can}^{\rm B} \hat H^{\rm B})]$, where we assume that the state of S changes adiabatically during the measurement, that is, S does not directly exchange heat with M or B during the measurement. In other words, we regard the direct energy flows between M and S as work.  
We have also assumed that there is no direct energy flow between  S and  B. 

To relate the information gain from the measurement to its thermodynamic energy cost, we  introduce  the Shannon information $H  \equiv - \sum_k p_k \ln p_k$ of the measurement outcomes and the QC-mutual information  between S and M, $I \equiv S(\hat \rho_{\rm i}^{\rm S}) +  H + \sum_k {\rm tr}[\sqrt{\hat E}_k \hat \rho_{\rm i}^{\rm S} \sqrt{\hat E_k} \ln \sqrt{\hat E_k} \hat \rho_{\rm i}^{\rm S} \sqrt{\hat E_k}]$~\cite{Sagawa-Ueda}, where $S(\hat \rho) \equiv -{\rm tr}(\hat \rho \ln \hat \rho)$ is the von Neumann entropy.
The QC-mutual information $I$ characterizes an effective information that is obtained by  quantum measurement and satisfies $0 \leq I \leq H$, where $I=H$   if the measurement is error-free and classical and $I=0$  if  no information is obtained from the measurement.
The QC-mutual information reduces to the classical mutual information~\cite{Cover-Thomas} in the classical limit.   

The first main result of this study is the lower bound on the work $W_{\rm meas}^{\rm M}$ required for the measurement:
\begin{equation}
W_{\rm meas}^{\rm M} \geq -k_{\rm B}T ( H - I) + \Delta F^{\rm M}.
\label{meas}
\end{equation}
The proof of this inequality is given later.
We note that $H-I$ satisfies $0 \leq H-I \leq H$; 
the lower bound on the work required increases as the amount of information gain $I$ by the measurement increases. 
The fundamental thermodynamic energy cost of measurement can be determined from inequality (\ref{meas}), regardless of  the state of the measured system S.  For the special case where $I=H$ and $\Delta F^{\rm M} = 0$, the right-hand side of (\ref{meas}) vanishes; this is in agreement with the fact that  there is no fundamental energy cost for  measurement and  communication~\cite{Landauer,Bennett2}.  
The lower bound of the work for given information contents ($H$ and $I$) and thermodynamic constraint ($\Delta F^{\rm M}$) is established by inequality (\ref{meas}).

We now discuss the thermodynamic energy cost of the erasure of information obtained by the measurement.  We treat M+B as an isolated quantum system. 
Suppose the initial state of M obeys the canonical distribution such that the probability of measurement outcome ``$k$'' is $p_k$. The initial state of M and B is described as $\hat \rho_{\rm i}^{\rm MB} = \sum_k p_k \hat \rho_{k, {\rm can}}^{\rm M} \otimes \hat \rho_{\rm can}^{\rm B}$, where $\hat \rho_{k, {\rm can}}^{\rm M} \equiv \exp (- \beta \hat H^{\rm M}_k)/Z_k^{\rm M}$.
The total system  evolves unitarily, and the support of the  final density operator, from the definition of information erasure, belongs to the subspace corresponding to the standard state $k=0$  with unit probability.
Let  $\hat \rho^{\rm BM}$  be the density operator of the final state of B and M.
The work required for the erasure is defined as $W_{\rm eras}^{\rm M} \equiv  [{\rm tr} (\hat \rho^{\rm MB}\hat H_0^{\rm M}) + {\rm tr}(\hat \rho^{\rm MB} \hat H^{\rm B})] 
 - \sum_k p_k [ {\rm tr} (\hat \rho_{k, {\rm can}}^{\rm M} \hat H_k^{\rm M}) + {\rm tr}(\hat \rho_{\rm can}^{\rm B}\hat H^{\rm B} ) ]$.
The lower bound on $W_{\rm eras}^{\rm M}$ is given by
\begin{equation}
W_{\rm eras}^{\rm M} \geq k_{\rm B} T H - \Delta F^{\rm M},
\label{eras}
\end{equation}
which is the second main result of this study. The proof of this inequality is given  later.
For the special case in which $F_0^{\rm M} = F_k^{\rm M}$ for all $k$, and hence $\Delta F^{\rm M} = 0$, we obtain  $W_{\rm eras} \geq k_{\rm B} T H$; this is in agreement with Landauer's principle~\cite{Landauer,Piechocinska}.  
However, when $\Delta F^{\rm M}\neq 0$, information erasure with $W_{\rm eras}^{\rm M} < k_{\rm B} T H$, in particular, with $W^{\rm M}_{\rm eras}=0$ is possible. 
Thus, there is no fundamental energy cost of information erasure as in the case of  measurement.
An inequality similar but not equivalent to (\ref{eras}) has recently been derived in Ref.~\cite{Maroney}.

Combining (\ref{meas}) and (\ref{eras}), we obtain
\begin{equation}
W^{\rm M}_{\rm meas} + W^{\rm M}_{\rm eras} \geq k_{\rm B} T I.
\label{Energy-cost2}
\end{equation}
This inequality shows that the lower bound on the total thermodynamic energy cost of measurement and information erasure depends neither on the Shannon information content nor on the free-energy difference; rather the bound depends only on the mutual information content between the measured system and the  memory.  
Inequality (\ref{Energy-cost2}) expresses the trade-off between the work required for erasure and that required for measurement.
If the work required for erasure is negative, the work required for measurement must be positive, and vice versa.  
Although there is no fundamental lower bound on the  work required only for measurement or only for erasure, there exists a fundamental lower bound on their sum. 
This trade-off can be confirmed by considering the special model discussed below.
Note that in the case of reversible measurement with $W^{\rm M}_{\rm meas} = 0$, inequality~(\ref{Energy-cost2}) reduces to $W^{\rm M}_{\rm eras} \geq k_{\rm B}T H$. 
While we have adopted the commonly used definitions for measurement and erasure~\cite{Demon},  $W^{\rm M}_{\rm meas}$ and $W^{\rm M}_{\rm eras}$ can, of course, change if we choose different definitions.  However,  the crucial fact here is that for given definitions of $W^{\rm M}_{\rm meas}$ and $W^{\rm M}_{\rm eras}$,  we can still change their ratio by changing the physical structure of the memory.

Inequalities (\ref{meas}), (\ref{eras}), and (\ref{Energy-cost2}) constitute  the second law of ``information thermodynamics,'' in which information content and  thermodynamic variables are treated on an equal footing.
In the limit of $H \to 0$ and $I \to 0$, these inequalities are equivalent to the conventional second law of thermodynamics.

We now discuss the consistency of our results with the second law of thermodynamics.  
Recently, we have identified the upper bound of  work that can be extracted from a heat bath at temperature $T$ with the assistance of feedback control by  ``Maxwell's demon''~\cite{Sagawa-Ueda}: $W^{\rm S}_{\rm ext} \leq - \Delta F^{\rm S} + k_{\rm B}TI$,
where $W^{\rm S}_{\rm ext} \equiv - W^{\rm S}$ is the work extracted by the demon and $\Delta F^{\rm S}$ is the free-energy difference of the controlled system. This upper bound is larger than that of the conventional second law of thermodynamics by $k_{\rm B}TI$.  
Adding this inequality to (\ref{Energy-cost2}), we obtain
\begin{equation}
W^{\rm SM}_{\rm ext} \equiv  W^{\rm S}_{\rm ext} - W^{\rm M}_{\rm meas} - W^{\rm M}_{\rm er}  \leq - \Delta F^{\rm S},
\label{Reconciliation2}
\end{equation}
which implies that the conventional second law of thermodynamics is applicable for the entire system of the measured system and the demon.

As an illustration, we construct a model of memory in which information can be erased without work. 
The model includes a Brownian particle moving in a double-well potential (upper row in FIG. 1 (a)-(c))~\cite{Landauer,Bennett2}.
The particle is in the left (right) well when the memory registers ``$0$'' (``$1$'').  We assume that the height of the potential barrier far exceeds  both quantum and thermal fluctuations, so that the barrier is impenetrable, and that the potential can be modeled by two boxes (lower row in  FIG. 1 (a)-(c)) with  the volume ratio being   $t:1-t$ ($0 < t < 1$). 
We also assume that the double-well potential can deform into a single-well potential during the measurement and erasure. 
We note that the model illustrated in FIG.1 is not applicable to a measured system such as the Szilard engine~\cite{Szilard}; rather it is only applicable to the memory that stores the measurement outcome using the representation of a single-molecule gas.

\begin{figure}[htbp]
 \begin{center}
  \includegraphics[width=70mm]{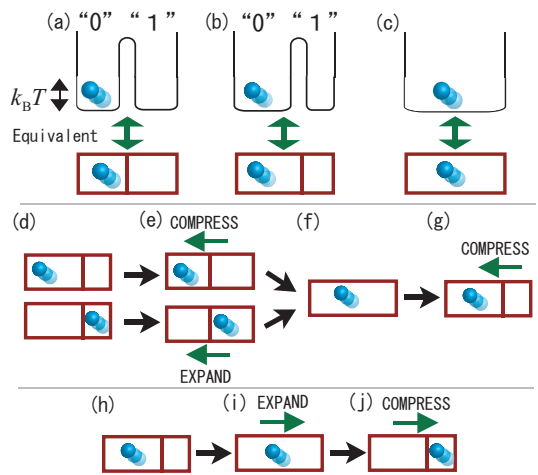}
 \end{center}
 \caption{(Color online) (a)-(c): Memory model that includes a Brownian particle moving in a double-well potential (upper row) as well as two boxes (lower row), where (a), (b), and (c) illustrate symmetric, asymmetric, and single potentials, respectively;
(d)-(g): information erasure from an asymmetric memory; 
(h)-(j): the state evolution of the memory for the case of outcome  ``$1$.''
}
\end{figure}

Let  the initial probabilities of obtaining outcomes ``$0$'' and ``$1$'' both be equal to $1/2$.
We consider a quasi-static information erasure at temperature $T$ as shown in  FIG. 1 (d)-(g).  
First, the memory stores the information concerning the outcome of the measurement [FIG.1(d)].
The partition is then moved to the center at an average work cost of  $(k_{\rm B}T/2) [\ln 2t +  \ln 2(1-t)]$ [FIG.1(e)]. The partition is then removed [FIG.1(f)].  This removal can be regarded as the free expansion of the gas, and therefore no work is required for the removal.
The box is finally  compressed at a work cost of  $-k_{\rm B}T \ln t$, and the memory returns to the standard state ``$0$'' [FIG.1(g)].
The total work required for information erasure is
$W_{\rm eras}^{\rm M}= k_{\rm B}T \ln2 - (k_{\rm B}T/2)\ln(t/(1-t))$. 
For the special case where $t=1/2$ (symmetric potential),  $W_{\rm eras}^{\rm M}= k_{\rm B}T \ln2$, as in the case of Bennett's model~\cite{Bennett2}.
In contrast, for $t=4/5$, $W_{\rm eras}^{\rm M}=0$ and no work is required for information erasure.
In general, $W_{\rm eras}^{\rm M} < k_{\rm B}T \ln2$ holds for $t>1/2$.  
Landauer's principle for information erasure is  valid for a symmetric double-well potential, but not for an asymmetric one.
The proof of $W_{\rm eras}^{\rm M} \geq k_{\rm B}T \ln 2$ using statistical mechanics in Ref.~\cite{Piechocinska} is valid only for the symmetric case.
We note that an asymmetric memory has also been discussed in Ref.\cite{Barkeshli}. 
We also note that for the case of quasi-static information processing for a given outcome (``$0$'' or ``$1$''), the ensemble average of the work always equals the work performed on an individual sample.  
In any case, we must average the work over all measurement outcomes.

We next consider a quasi-static measurement process at temperature $T$.
At the initial stage, the memory is in the standard state ``$0$.''
If the measurement outcome is ``$0$,'' the state of the memory does not change.
If the measurement outcome is ``$1$,''  the memory  interacts with the measured system, and the left box of the memory expands to the right; this requires  $-k_{\rm B} T \ln (1/t)$ of work [FIG. 1 (i)].  The box then compresses from the left at a work cost of $k_{\rm B} T \ln (1/(1-t))$ until the volume of the right box returns to the initial volume [FIG. 1 (j)].
The total work required in this case is given by $k_{\rm B} T \ln (t/(1-t))$.
Averaging the work over the measurement outcomes, we find that $W_{\rm meas}^{\rm M} = (k_{\rm B}T/2)\ln (t/(1-t))$  is required for the measurement.
Adding this to $W_{\rm eras}^{\rm M}= k_{\rm B}T \ln2 - (k_{\rm B}T/2)\ln(t/(1-t))$,  we find that  the total work required for measurement and erasure is $W_{\rm meas}^{\rm M} + W_{\rm eras}^{\rm M} = k_{\rm B}T \ln2$; 
the equality in (\ref{Energy-cost2}) can be attained by using this model.

More generally, let us consider the entropy balance of information erasure.  If the density operators of the memory, which we denote by $\hat \rho_k$'s, are mutually orthogonal, then the total entropy of $\hat \rho \equiv \sum_k p_k \hat \rho_k$ satisfies $S(\hat \rho ) = H + \sum_k p_k S(\hat \rho_k)$, where $H$ is the Shannon information and the $S(\hat \rho_k)$'s describe the physical entropy of the memory.  If the $S(\hat \rho_k)$'s are equal, then $\sum_k p_k S(\hat \rho_k)$ is independent of $\{ p_k \}$, and therefore, a decrease in $H$ must be compensated for by an increase in the entropy of an external heat bath due to the unitarity of the entire system.
If the $S(\hat \rho_k)$'s are not equal, a decrease in $H$ can be compensated for by a change in the physical entropy of the memory, $\sum_k p_k S(\hat \rho_k)$.
We illustrate this fact by using our model.  Let $V$ be the volume of the box. The physical entropy of the initial and final states are  $[ \ln (tV) + \ln ((1-t)V) ] / 2$  and $\ln (tV)$, respectively. The Shannon information that the memory stores in the initial state is $\ln 2$, which reduces to $0$ during the erasure process.
For the case where $t=1/2$, the difference between the total entropy in the initial and final states is $- \ln 2$, which must  eventually be dissipated into the heat bath.
On the other hand, for the case where $t=4/5$, the total entropy does not change during the erasure process.  In this case, a decrease in the Shannon information can be compensated for by an increase in the physical entropy of the memory itself.

We now prove inequality  (\ref{meas}).  We first prove  Eq.~(\ref{assumption}) for a classical case, that is, the case where all density operators are diagonal at all times with respect to an eigenbasis set corresponding to the classical degrees of freedom.   Let $\{ | ki \rangle \}_{k,i}$ be the eigenbasis set of M+B.  For a classical case, we can write $\hat \rho'^{\rm SMB} = \sum_{k, i}  \hat \rho^{\rm S}_{ki}   \otimes |  ki \rangle \langle ki |$, where $\hat \rho^{\rm S}_{ki}$'s are not normalized.  Our task is to show that there exists some $\hat M_k$ such that $\hat \rho^{\rm S}_{ki} = \hat M_{ki} \hat \rho_{\rm i}^{\rm S} \hat M_{ki}^\dagger$. Let $\hat \rho_{\rm i}^{\rm S} \equiv \sum_s q_s | s \rangle \langle s |$ and  $\hat \rho_{0,{\rm can}}^{\rm M} \otimes \hat \rho_{\rm can}^{\rm B} \equiv \sum_{k, i}r_{ki} | ki \rangle \langle ki |$.  Then, 
$\langle s | \rho^{\rm S}_{ki} | s'' \rangle =  \delta_{s,s''} \sum_{s', l, j} r_{ki}   \langle ski | U_{\rm int} | s'lj \rangle q_{s'} \langle s'lj | U_{\rm int}^\dagger  | s''ki \rangle$, where $\delta_{s,s''}$ is the Kronecker delta. Since without loss of generality, the unitary matrix of $\hat U_{\rm int}$ in the classical eigenbasis can be represented by a permutation matrix for a classical case,   for a given $(s, k, i)$, $\langle s'lj | U_{\rm int}^\dagger | ski \rangle$ is unity for one and only one  $(s', l, j)$ and is zero otherwise. We then have   $\langle s | \rho^{\rm S}_{ki} | s \rangle = \sum_{s'ljl'j'} r_{ki}   \langle ski | U_{\rm int} | s'lj \rangle q_{s'} \langle s'l'j' | U_{\rm int}^\dagger  | ski \rangle$.  By defining $\hat M_{ki} \equiv \sum_{l, j} \sqrt{r_{ki}} \langle ki | U_{\rm int} | lj \rangle$, we obtain Eq.~(\ref{assumption}).

Since the time evolution from  $\hat \rho_{\rm i}^{\rm SMB}$ to $\hat \rho_{\rm f}^{\rm SMB}$ is composed of the unitary evolution and the projection, we have $S(\hat \rho_{\rm i}^{\rm SMB}) \leq S(\hat \rho_{\rm f}^{\rm SMB})$.  On the other hand, we can show that $
S (\hat \rho_{\rm f}^{\rm SMB} ) =  H  + \sum_k p_k  S(\sum_{i}   \hat M_{ki} \hat \rho_{\rm i}^{\rm S} \hat M_{ki}^\dagger   \otimes \hat \rho_{i}^{\rm MB}/p_k)$ and that $S(\sum_{i}   \hat M_{ki} \hat \rho_{\rm i}^{\rm S} \hat M_{ki}^\dagger   \otimes \hat \rho_{ki}^{\rm MB}/p_k) = \sum_{i} (p_{ki}/p_k) S(\hat M_{ki} \hat \rho_{\rm i}^{\rm S} \hat M_{ki}^\dagger / p_{ki}) + S(\hat \rho_k^{\rm MB})$.  Noting that $S(\hat M_{ki} \hat \rho_{\rm i}^{\rm S} \hat M_{ki}^\dagger) = S(\sqrt{\hat \rho_{\rm i}^{\rm S}} \hat M_{ki}^\dagger \hat M_{ki} \sqrt{\hat \rho_{\rm i}^{\rm S}})$ holds and that the von Neumann entropy is concave,  we can show that $\sum_{i} (p_{ki}/p_k) S(\hat M_{ki} \hat \rho_{\rm i}^{\rm S} \hat M_{ki}^\dagger / p_{ki}) \leq S(\sqrt{\hat E_k} \hat \rho_{\rm i}^{\rm S} \sqrt{\hat E_k}/p_k)$.
From the definition of the QC-mutual information content $I$, we obtain $\sum_k p_k S(\hat \rho_k^{\rm MB}) - S(\hat \rho_{0, {\rm can}}^{\rm M}) - S(\hat \rho_{\rm can}^{\rm B}) \geq I - H$.
It follows from  Klein's inequality that $- \sum_k p_k {\rm tr} [\hat \rho_k^{\rm MB}  \ln \hat \rho_{k, {\rm can}}^{\rm M} \otimes \hat \rho_{\rm can}^{\rm B}]  - S(\hat \rho_{0, {\rm can}}^{\rm M}) - S(\hat \rho_{\rm can}^{\rm B}) \geq I - H$.  From the definition of the work, 
we finally obtain $- \Delta F^{\rm M} + W_{\rm meas}^{\rm M} \geq k_{\rm B}T (I - H)$, which proves inequality~(\ref{meas}).

We next  prove inequality (\ref{eras}).    Noting that the evolution of $\hat \rho^{\rm BM}$ is unitary, we have $S(\hat \rho^{\rm MB}) - \sum_k p_k S(\hat \rho_{k, {\rm can}}^{\rm M}) - S(\hat \rho_{\rm can}^{\rm B}) = H$.
From Klein's inequality, we have $
- {\rm tr} (\hat \rho^{\rm MB} \ln \hat \rho_{0, {\rm can}}^{\rm M} \otimes \hat \rho_{\rm can}^{\rm B}) - \sum_k p_k S(\hat \rho_{k, {\rm can}}^{\rm M}) - S(\hat \rho_{\rm can}^{\rm B}) \geq H$, which leads to inequality~(\ref{eras}).

Eighty years ago Szilard discovered the close relationship between information and thermodynamics and suggested that ``it will be possible to find a more general entropy law, which applies universally to all measurements"~\cite{Szilard}.  
Since Szilard's discovery, this crucial insight has been expanded and deepened.
In 1951, on the basis of a specific model, Brillouin argued that the work is needed for the measurement, which compensates for the excess work extracted by Maxwell's demon~\cite{Brillouin}.  Later, Bennett proposed a model of the demon that can perform measurement without any work and on the basis of a specific model and Landauer's principle argued that the key to resolving the paradox of the demon lies in erasing the information stored in the demon's memory~\cite{Bennett2}.  Since then, it has been widely believed that the work required for information erasure compensates for the excess work~\cite{Demon}.  In this Letter, we have derived inequalities~(\ref{Energy-cost2}) and (\ref{Reconciliation2}) that unify the approach adopted by Szilard, Brillouin, and Bennett, and we have shown that 
what reconciles Maxwell's demon with the second law of thermodynamics is the total work of the measurement and erasure, which compensates for the excess work of $k_{\rm B}TI$ that can be extracted by the demon.

\begin{acknowledgments}
This work was supported by a Grant-in-Aid for Scientific Research (Grant No.\ 17071005), and by a Global COE program ``Physical Science Frontier'' of MEXT, Japan. TS acknowledges JSPS Research Fellowships for Young Scientists (Grant No. 208038).
\end{acknowledgments}

\end{document}